\documentclass[11pt]{article}
\usepackage{graphicx}

\newcommand{\BABARPubYear}    {06}

\newcommand{\BABARConfNumber} {011}
\newcommand{\SLACPubNumber} {12008}

\input babarsym

\long\def\inst#1{\par\nobreak\kern 4pt\nobreak
    {\it #1}\par\vskip 10pt plus 3pt minus 3pt}

\RequirePackage{xspace}

\newcommand{\pvec}{{\bf p}}

\newcommand{\calB}{\ensuremath{{\cal B}}}

\newcommand{\DE}{\ensuremath{\Delta E}}

\newcommand{\xf}{\ensuremath{{\cal F}}}

\newcommand{\thetaT}{\ensuremath{\theta_{\rm T}}}
\newcommand{\costhr}{\ensuremath{\cos\thetaT}}

\newcommand\etal{{\it et al.}}
\newcommand{\half}{\ensuremath{{1\over2}}}

\newcommand{\bfig}{\begin{figure}[htbpc!]}
\newcommand{\efig}{\end{figure}}
\newcommand\bef{\begin{figure}}
\newcommand\edf{\end{figure}}
\newcommand\dbline{\noalign{\vskip 0.10truecm\hrule}\noalign{\vskip 2pt}\noalign{\hrule\vskip 0.10truecm}}

\newcommand\sgline{\noalign{\vskip 0.10truecm\hrule\vskip 0.10truecm}}
\newcommand\beq{\begin{equation}}
\newcommand\eeq{\end{equation}}
\newcommand\bear{\begin{array}}
\newcommand\enar{\end{array}}
\newcommand\beqa{\begin{eqnarray}}
\newcommand\eeqa{\end{eqnarray}}
\newcommand\ben{\begin{enumerate}}
\newcommand\een{\end{enumerate}}

\newcommand{\UfourS}{\ensuremath{\Upsilon(4S)}}

\newcommand{\kstopipi}{\ensuremath{\KS\to\pi^+\pi^-}}

\newcommand{\bsg}{\ensuremath{b\rightarrow s g^{\star}}}
\newcommand{\BKX}{\ensuremath{B\rightarrow KX}}

\newcommand{\BKpX}{\ensuremath{B\rightarrow K^+ X}}
\newcommand{\BKsX}{\ensuremath{B\rightarrow K_S^0 X}}
\newcommand{\BKzX}{\ensuremath{B\rightarrow K^{0} X}}
\newcommand{\btou}{\ensuremath{b\rightarrow u}}
\newcommand{\btod}{\ensuremath{b\rightarrow d}}
\newcommand{\btos}{\ensuremath{b\rightarrow s}}
\newcommand{\btoc}{\ensuremath{b\rightarrow c}}

\newcommand{\pstar}{\ensuremath{p^{\star}}}
\newcommand{\pstarK}{\ensuremath{\pstar(K)}}
\newcommand{\pstarKp}{\ensuremath{\pstar(K^+)}}

\newcommand{\pstarKs}{\ensuremath{\pstar(K_S^0)}}
\newcommand{\pstarKz}{\ensuremath{\pstar(\Kz)}}

\newcommand{\bsig}{\ensuremath{B_{\rm signal}}}
\newcommand{\breco}{\ensuremath{B_{\rm reco}}}

  \newcommand{\pBRBKpX}{\ensuremath{196^{+37}_{-34}(\rm stat.)^{+31}_{-30}(\rm syst.)}}        
  \newcommand{\VpBRBKpX}{\ensuremath{196^{+37}_{-34}{}^{+31}_{-30}}}        
  \newcommand{\signifBKpX}{\ensuremath{6.0}}                     

  \newcommand{\pbrBKpX}{\ensuremath{\calB(\BKpX, \pstar>2.34\gev)}}

  \newcommand{\pBRBKsX}{\ensuremath{154^{+55}_{-48}(\rm stat.)^{+55}_{-41}(\rm syst.)}}        
  \newcommand{\VpBRBKsX}{\ensuremath{154^{+55}_{-48}{}^{+55}_{-41}}}        
  \newcommand{\ulpBRBKsX}{\ensuremath{266}}                    
  \newcommand{\signifBKsX}{\ensuremath{3.1}}                     

  \newcommand{\pbrBKzX}{\ensuremath{\calB(\BKzX, \pstar>2.34\gev)}}
  \newcommand{\brBKzX}{\ensuremath{\calB(\BKzX)}}

\begin{document}
{\pagestyle{empty}

\begin{flushright}
\babar-CONF-\BABARPubYear/\BABARConfNumber \\
SLAC-PUB-\SLACPubNumber \\
\end{flushright}

\par\vskip 5cm

\begin{center}
\Large \bf Search for inclusive charmless \boldmath{\BKpX} and \boldmath{\BKzX} decays
\end{center}
\bigskip

\begin{center}
\large The \babar\ Collaboration\\
\mbox{ }\\
\today
\end{center}
\bigskip \bigskip

\begin{center}
\large \bf Abstract
\end{center}
We present
preliminary results from
a search for inclusive charmless \BKX\ decays.
These decays occur dominantly via one-loop \btos\ penguin
transitions, and can provide useful information about
these processes. Using a sample of $288.5$~\invfb\
collected with the \babar\ detector at the PEP-II
asymmetric-energy \epem\ $B$ Factory at SLAC, we search
for high-energy kaons recoiling against fully
reconstructed $B$ decays. We measure the partial branching
fractions for kaons with momentum $\pstarK>2.34$\gev\ in the
$B$ rest frame, and obtain
(in units of $10^{-6}$): $\pbrBKpX = \pBRBKpX$ and
$\pbrBKzX = \pBRBKsX$ ($<\ulpBRBKsX$ at $90\%$ C.L.).

\vfill
\begin{center}

Submitted to the 33$^{\rm rd}$ International Conference on High-Energy Physics, ICHEP 06,\\
26 July---2 August 2006, Moscow, Russia.

\end{center}

\vspace{1.0cm}
\begin{center}
{\em Stanford Linear Accelerator Center, Stanford University, 
Stanford, CA 94309} \\ \vspace{0.1cm}\hrule\vspace{0.1cm}
Work supported in part by Department of Energy contract DE-AC03-76SF00515.
\end{center}

\newpage
} 

%
\begin{center}
\small

The \babar\ Collaboration,
\bigskip

%
{B.~Aubert,}
{R.~Barate,}
{M.~Bona,}
{D.~Boutigny,}
{F.~Couderc,}
{Y.~Karyotakis,}
{J.~P.~Lees,}
{V.~Poireau,}
{V.~Tisserand,}
{A.~Zghiche}
\inst{Laboratoire de Physique des Particules, IN2P3/CNRS et Universit\'e de Savoie,
 F-74941 Annecy-Le-Vieux, France }
% ...

\end{center}\newpage

\section{INTRODUCTION}
\label{sec:Introduction}
$B$-meson decays proceed dominantly through \btoc\ transitions, while
the tree-level \btou\ and one-loop \btos\ transitions
are suppressed.
In the Standard Model (SM), 
the branching fractions for \btou\ and \btos\ transitions are approximately
$1\%-2\%$~\cite{HSS87,GL01,LNO97}.
It has been suggested that loop transitions are a window on the effects
of new physics, as virtual non-SM particles in the
loop can couple to the quarks~\cite{BBSV93,GIS01}.
The branching fraction for \bsg\ ($g^{\star}=$ gluon)
decays could be as large as $10\%$ in certain models~\cite{BBSV93,GIS01}.

In recent years, exclusive $B$ decays dominated by \bsg\
($b\rightarrow sq\overline{q}$,
$q=u,d,s$) have been used to measure the CKM unitarity triangle angle
$\beta$. The amplitude $S$ of the sine component of the time-dependent $CP$ asymmetry
in these decay modes is measured to be systematically shifted low relative to
the expected values from SM calculations ($S\simeq\sin 2\beta$),
although this shift is currently not statistically significant~\cite{HFAG06}.

A good understanding of the dynamics of
\btos\ transitions is needed to make accurate predictions of related quantities
within the framework of the SM~\cite{BHNR05}.
This understanding currently comes from
branching fractions and $CP$ measurements of \btos\ dominated exclusive
decays and from inclusive and exclusive $b\rightarrow s\gamma$ transitions. The
measurement of inclusive \bsg\ decays would provide additional
information to the current picture~\cite{HK04}, and could help us
understand the discrepancies seen in the measurements of
$\sin 2\beta$~\cite{BHNR05}.

Previous experimental attempts to measure inclusive \bsg\ decays have
been statistically limited~\cite{ARGUS95,CLEO98,DELPHI98,SLD98}.
The $B$-factory experiments present new opportunities to make a significant
measurement of this process.

In this paper, we present a preliminary result from a search for inclusive charmless \BKpX\ and
\BKzX\ decays, which can in principle be related to the \bsg\ rate.
The neutral kaon in \BKzX\ is reconstructed through the decay $K_S^0\rightarrow\pi^+\pi^-$.
We define as signal \BKX\ all the charmless decays that contain at least
one kaon. These decays can occur via \btos\ (dominant), \btou, and \btod\
transitions.
The signal yields are extracted with an unbinned maximum likelihood (ML)
fit to samples of \BKX\ decays recoiling against fully reconstructed
hadronic $B$ decays.

\section{THE \babar\ DETECTOR AND DATASET}
\label{sec:babar}
The data used in this analysis were collected with the \babar\
detector~\cite{ref:babar} at the \pep2\ asymmetric-energy \epem\
collider located at the Stanford Linear Accelerator Center (SLAC). The
analysis uses an integrated luminosity of $288.5$\invfb\ recorded
at the $\Upsilon (4S)$ resonance (center-of-mass energy
$\sqrt{s}=10.58\ \gev$).

In the \babar\ detector, charged particles are detected and their
momenta measured by a combination of a vertex tracker consisting
of five layers of double-sided silicon microstrip detectors and a
40-layer central drift chamber, both operating in the 1.5-T magnetic
field of a superconducting solenoid. We identify photons and electrons 
using a CsI(Tl) electromagnetic calorimeter (EMC).
Charged particle identification (PID) is provided by
an internally reflecting ring imaging Cherenkov detector (DIRC)
covering the central region of the detector,
the average energy loss ($dE/dx$) in the tracking devices and by
the EMC. Additional information that we use to identify and reject
electrons and muons is provided by the EMC and the detectors
of the solenoid flux return (IFR).

\section{ANALYSIS METHOD AND EVENT SELECTION}
\label{sec:Analysis}
The experimental identification of the inclusive \bsg\ decay is complicated by the
fact that the gluon is a virtual intermediate state with no good experimental
signature. We instead rely on the hadronization of the primary strange quark
into a charged or neutral kaon to identify decays dominated by \btos\ transitions.
The analysis is therefore effectively a search for the decays \BKpX\ and \BKzX .
The momentum \pstarK\ of the primary
kaon in the $B$ rest frame is limited for \btoc\ background by the $D$-meson mass,
and cannot be larger than $\sim 2.3$\gev, while \pstarK\ can be as large as
$\sim 2.6$\gev\ for signal \BKX\ decays. We use this difference as the primary
signature to look for charmless inclusive \BKX\ decays, as was first
suggested in Ref.~\cite{BDHP98}.

In this analysis we reject  the large $e^+e^-\rightarrow q\overline{q}$ ($q=u,d,s,c$)
continuum background very efficiently by reconstructing one of the two $B$ mesons (\breco)
in $e^+e^-\rightarrow\Upsilon(4S)\rightarrow B\overline{B}$, and searching for
the \BKX\ signal (\bsig) recoiling against \breco.
We select a large sample of events containing a \breco\ meson
which decays into a hadronic final state as
$\breco\rightarrow\overline{D}^{(*)}Y^{\pm}$, and is fully reconstructed.
The system $Y^{\pm}$ consists of a combination of hadrons containing
one, three, or five charged pions or kaons, up to two neutral pions,
and at most two $\KS\rightarrow\pi^+\pi^-$ candidates. We reconstruct
$D^{* -}\rightarrow \overline{D}^{0}\pi^{-}$;
$D^{* 0}\rightarrow \overline{D}^{0}\pi^{0}$;
$\overline{D}^0\rightarrow K^{+}\pi^{-}$, 
$K^{+}\pi^{-}\pi^{0}$, 
$K^{+}\pi^{-}\pi^{-}\pi^{+}$, 
$\KS\pi^{+}\pi^{-}$; and
$D^-\rightarrow K^{+}\pi^{-}\pi^{-}$, 
$K^{+}\pi^{-}\pi^{-}\pi^{0}$, 
$\KS\pi^{-}$, 
$\KS\pi^{-}\pi^{0}$,
$\KS\pi^{-}\pi^{-}\pi^{+}$.
The \breco\ candidates are characterized kinematically by the
energy-substituted mass
$\mes=(\frac{1}{4}s-\pvec_B^2)^\half$
and energy difference $\DE = E_B-\half\sqrt{s}$, where
$(E_B,\pvec_B)$ is the $B$-meson 4-momentum vector, and
all values are expressed in the \UfourS\ frame. We require
the value of \DE\ to be consistent with zero within three standard
deviations ($\sigma$), as measured for each decay mode ($10$ to $35$\mev). We
require $5.25<\mes<5.29$\gev, and use this variable in the ML fit
described below.
We define the purity for each \breco\ decay mode as the ratio
$S/(S+B)$ measured in control samples, where $S$ is the number of reconstructed signal \breco\
and $B$ is the number of background events. We require the
purity of the selected \breco\ candidates to be at least $20\%$.
In events containing more than one \breco\ candidate, we select the decay
mode with the highest purity.

To further reject \qqbar\ continuum
background, we make use of the angle \thetaT\ between the
thrust axis of the \breco\ candidate in the \UfourS\ frame and that of the 
rest of the charged tracks and neutral clusters in the event.
The distribution of $|\costhr|$ is
sharply peaked near $1$ for combinations drawn from jet-like \qqbar\
pairs, and nearly uniform for the almost isotropic $B$-meson decays;
we require $|\costhr|<0.9$. Further discrimination
from continuum in the ML\ fit is obtained from a Fisher discriminant \xf,
which is an optimized linear combination of four variables: the angles with 
respect to the beam axis of the $B$ momentum and $B$ thrust axis (in
the \UfourS\ frame), and the zeroth and second angular moments
$L_{0,2}$ of the energy flow about the $B$ thrust axis.  The moments
are defined by $ L_j = \sum_i p_i\times\left|\cos\theta_i\right|^j,$
where $\theta_i$ is the angle with respect to the $B$ thrust axis of
track or neutral cluster $i$, $p_i$ is its momentum, and the sum
excludes the \breco\ candidate.

We select $2.99\times10^6$ \breco\ candidates with the above criteria,
and apply an unbinned ML fit to the \mes\ and \xf\ variables to separate
\BB\ events from \qqbar\ continuum background. We find
$N(\breco)=(1.78\pm0.09)\times 10^6$, where the uncertainty includes a
conservative preliminary estimate of the systematics due to the modeling
of the data.

We search for the signal $B$ meson (\bsig) using the charged tracks and the neutral
clusters that are not part of the \breco\ candidate. We reject \bsig\ candidates
containing charged tracks compatible with an electron or muon hypothesis,
or that contain a reconstructed $D$ meson candidate with mass within $30$ \mev\
of the nominal mass. We also require $\mes(\bsig)>5.1$\gev .

The measured \UfourS\ and \breco\ 4-momentum vectors are used
to determine accurately the 4-momentum vector of \bsig, independently
of its decay products. We then select charged $K^+$ and neutral
$\KS\rightarrow\pi^+\pi^-$ candidates with momentum \pstarK\ larger than $2.34$\gev,
calculated in the \bsig\ rest frame.

The DIRC Cherenkov angle $\theta_c$ for charged kaon candidates must satisfy
$-5\sigma_c<\theta_{c}<+2\sigma_c$, where $\sigma_c$ is the resolution on $\theta_c$,
and the upper limit is designed to reject contamination from charged pions.
To exclude secondary kaons, the distance of
closest approach of the $K^+$ candidate must be within three standard
deviations of the \bsig\ vertex, as determined inclusively from all tracks
recoiling against \breco .

We require the mass of \KS\ candidates to be within $\pm4.5\sigma$ of the
nominal mass ($0.486<m_{\KS}<0.510$\gev). The reconstructed \kstopipi\ must have a vertex
$\chi^2$ probability ${\cal P}_{\rm vertex}>0.001$, and its lifetime significance
($\tau/\sigma_{\tau}$) must be larger than $3$.

The above selection has an efficiency of $(16.1\pm 1.5)\%$ for the decays
\BKpX\ with $\pstarKp>2.34$\gev, and $(6.7\pm1.1)\%$ for the decays \BKzX\
with $\pstarKz>2.34$\gev\ and reconstructed as \BKsX, $\KS\rightarrow\pi^+\pi^-$.
The selected samples contain $246$ \BKpX\ and $76$ \BKsX\ candidates.
We estimate from Monte Carlo simulation that $10-20\%$ of the selected
candidates come from \btoc\ decays, which can produce kaons of momentum
higher than $2.34$\gev\ in the decay of $D$ mesons. Contamination from
unflavored \btou\ and \btod\ transitions is negligible for \BKsX, and
estimated to contribute to $2.4\%$ of the \BKpX\ sample via $K/\pi$ mis-identification.

\section{MAXIMUM LIKELIHOOD FIT}
\label{sec:MLFit}
We obtain yields for each decay from an extended unbinned
ML fit with the following input observables: \mes, \xf, and \pstarK.
As described below, the fit is first applied with several probability
density function (PDF) parameters floating
to samples obtained with $\pstarK>1.8$\gev. The signal yield is then extracted
from a fit to the $\pstarK>2.34$\gev\ samples, in which the \BB\ background yield
and \pstarK\ PDF are fixed to the results of the first fit.

For each event $i$ and hypothesis $j$ (signal \BKX, \BB\ background,
continuum background), we define the  probability density function (PDF)
\begin{eqnarray}
{\cal P}^i_{j} =  {\cal P}_j (\mes^i) { \cal P}_j(\xf^i) {\cal P}_j (\pstar{}^i).
\end{eqnarray}
The likelihood function is
\begin{equation}
{\cal L} = \exp{(-\sum_{j} Y_{j})}
\prod_i^{N}\left[\sum_{j} Y_{j} {\cal P}^i_{j}\right]\,,
\end{equation}
where $Y_{j}$ is the yield of events of hypothesis $j$,
to be found by maximizing \calL . $N$ is the number of events in the sample.

\begin{figure}[ht]
\begin{center}
\includegraphics[width=.49\linewidth]{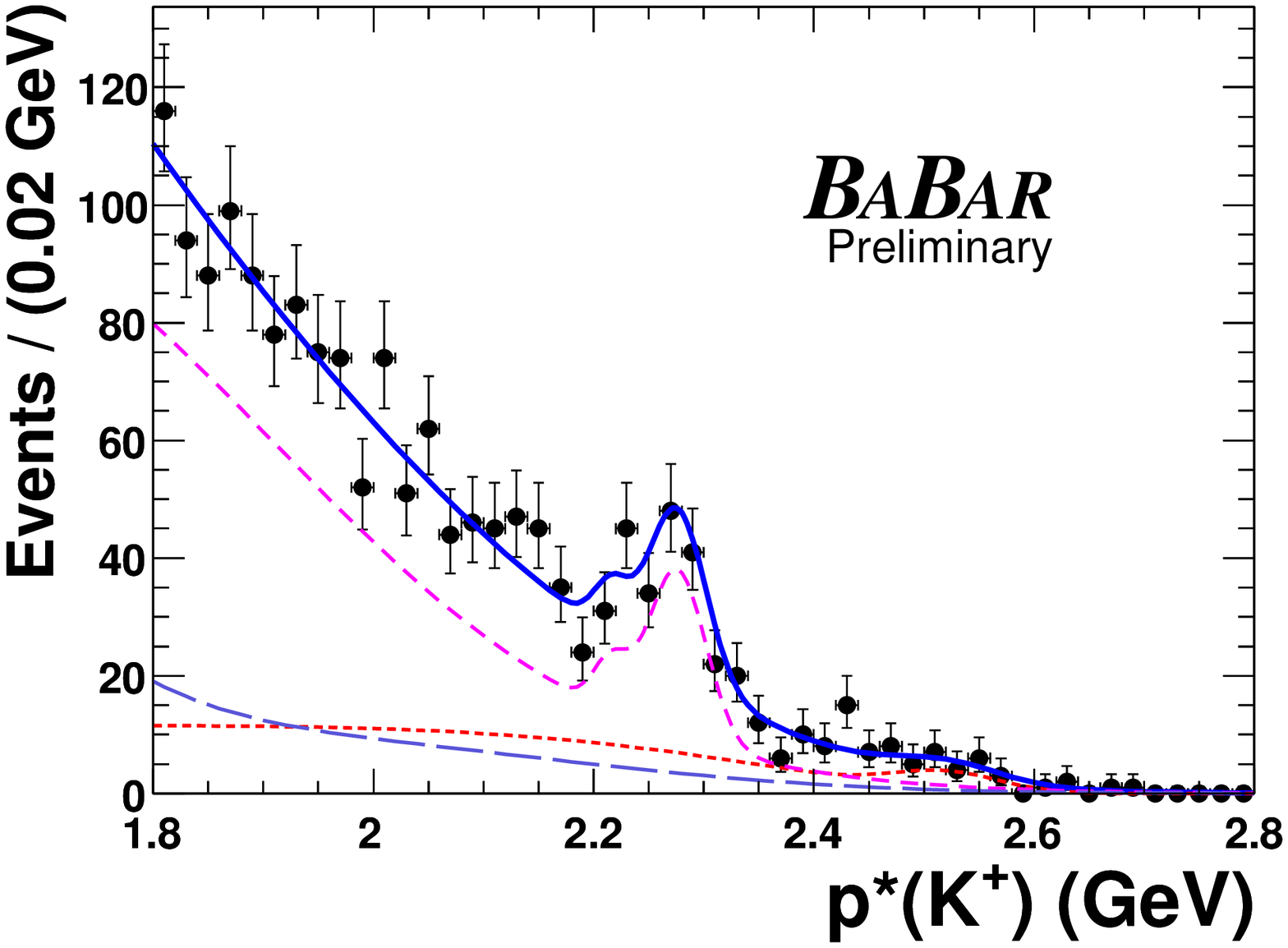}
\includegraphics[width=.49\linewidth]{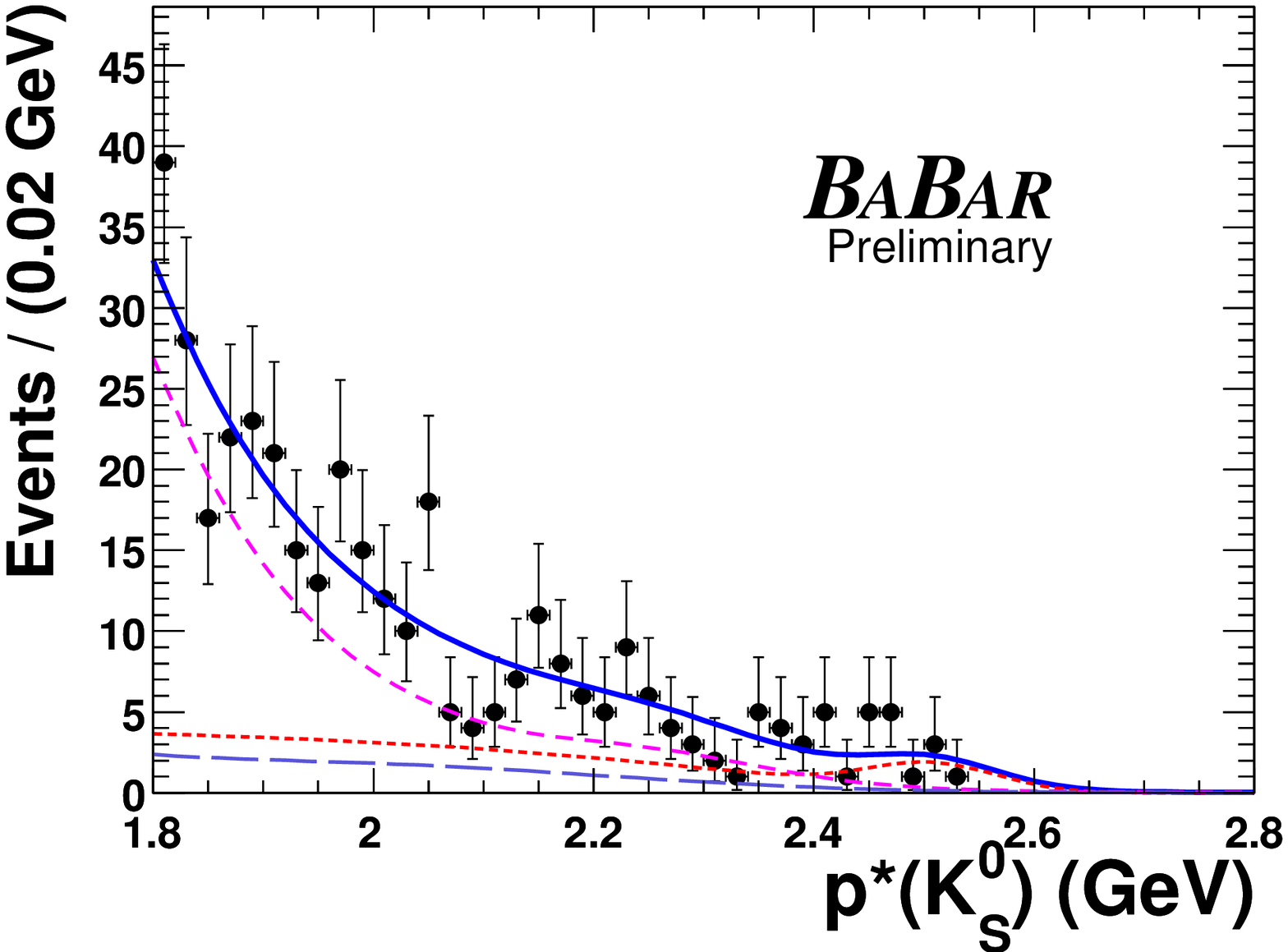}
\caption{Projection plots
for the \pstarKp\ (left) and \pstarKs\ (right) variables
from the fits to the $\pstarK>1.8$\gev\ samples.
The projections are obtained with a cut on the signal likelihood (see text)
retaining about $80\%$ of the signal events.
The points are
from the data, the full line shows the full fit,
the dotted line the signal,
the short-dashed line the \BB\ background,
and the long-dashed line the \qqbar\ continuum background.}
\label{fig:projPstarWide}
\end{center}
\end{figure}

The \mes\ and \xf\ variables discriminate between \BB\ and
\qqbar\ continuum events. For these variables, the same PDF is
used for the \BKX\ signal and \BB\ background components.
The \mes\ PDF for \qqbar\ continuum is parametrized by an empirical
phase-space function~\cite{ARGUS90} of the form
\begin{equation}
\label{eq:argus}
f(x)\propto x\sqrt{1-x^2}\exp{\left[-\xi(1-x^2)\right]}
\end{equation}
where $x\equiv 2\mes/\sqrt{s}$, and $\xi$ is a parameter determined 
by the fit.
For $B$ decays, \mes\ is modeled by the sum of two Gaussians and
the function of Eq.~\ref{eq:argus} with a different value of $\xi$.
The \xf\ PDF is parametrized as a bifurcated Gaussian plus a Gaussian
for \BB\ events, and as two Gaussians for \qqbar\ continuum.
The \pstarK\ PDF is defined over the wider range $\pstarK>1.8$\gev.
For the signal \BKX, the PDF is the sum of a phase-space function given
by Eq.~\ref{eq:argus}, with $x\equiv\pstarK/2.62$\gev, and a Gaussian to account for the
contribution from exclusive $2$-body decays such as
$B\rightarrow\eta^{\prime}K$. The parameters of the signal
\pstarK\ PDF will be varied in the evaluation of the systematic
errors, as the \pstarK\ spectrum is not well known.
The \BB\ background PDF is the sum
of three Gaussians, two of them used to model the
$B\rightarrow DK$ and $B\rightarrow D^{*}K$ contributions.
The \qqbar\ component is described by the sum of an exponential
and a Gaussian.
All the PDF distributions are illustrated in Figs.~\ref{fig:projPstarWide}
and~\ref{fig:ProjectionPlots}.

The PDF for each variable and each component is initially determined from
Monte Carlo (MC) simulation. A preliminary ML fit with several free PDF
parameters is applied to the sample obtained with the relaxed requirement
$\pstarK>1.8$\gev.
This range of \pstarK\ includes too much background for an accurate
determination of the signal yield, but it allows the measurement of the   
yield and the \pstarK\ PDF for the \BB\ component.
The free parameters of this fit are the three yields,
the fractions of neutral \breco\ candidates for each component,
the size of the \pstarK\ secondary Gaussians for the signal and \BB\ components,
the width of the \pstarK\ main Gaussian for the \BB\ component,
and the \mes\ exponent parameter for the \qqbar\ component.
The results are illustrated in
Fig.~\ref{fig:projPstarWide}, which shows the projections onto \pstarKp\ and \pstarKs\
of subsamples enriched with a threshold requirement on the signal likelihood
computed without the variable plotted.

The PDFs determined in the first fit are then used in a second ML fit
to the sample obtained with $\pstarK>2.34$\gev. Free parameters are the
signal and \qqbar\ continuum background yields, while the \BB\ yield is fixed
to the fraction of the value measured in the first fit ($\pstarK>1.8$\gev).
Systematic errors account for the uncertainties in the fixed \BB\ yield, as
determined in the first fit and which include the affect of correlations with the
signal component.

Monte Carlo simulated experiments are used to validate the fit procedure,
and to evaluate possible biases in the yields due to our neglect of
small residual correlations among discriminating variables.
The bias is determined by fitting ensembles of simulated \qqbar\ experiments
drawn from the PDF into which we have embedded the expected number of signal
and \BB\ background events, randomly extracted from the fully simulated MC
samples. The measured biases are listed in Table~\ref{tab:results}.

\section{RESULTS}
\label{sec:Physics}
\begin{figure}[ht]
\begin{center} 
\includegraphics[width=.8\linewidth]{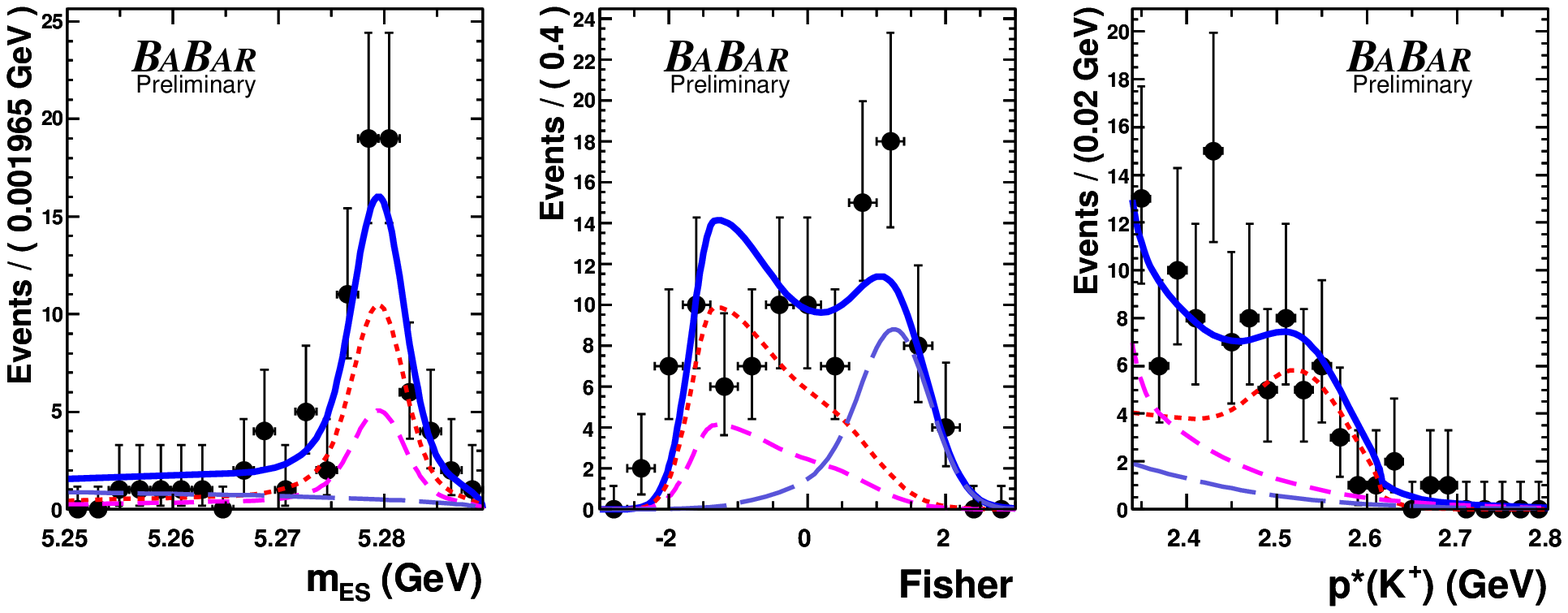} 
\includegraphics[width=.8\linewidth]{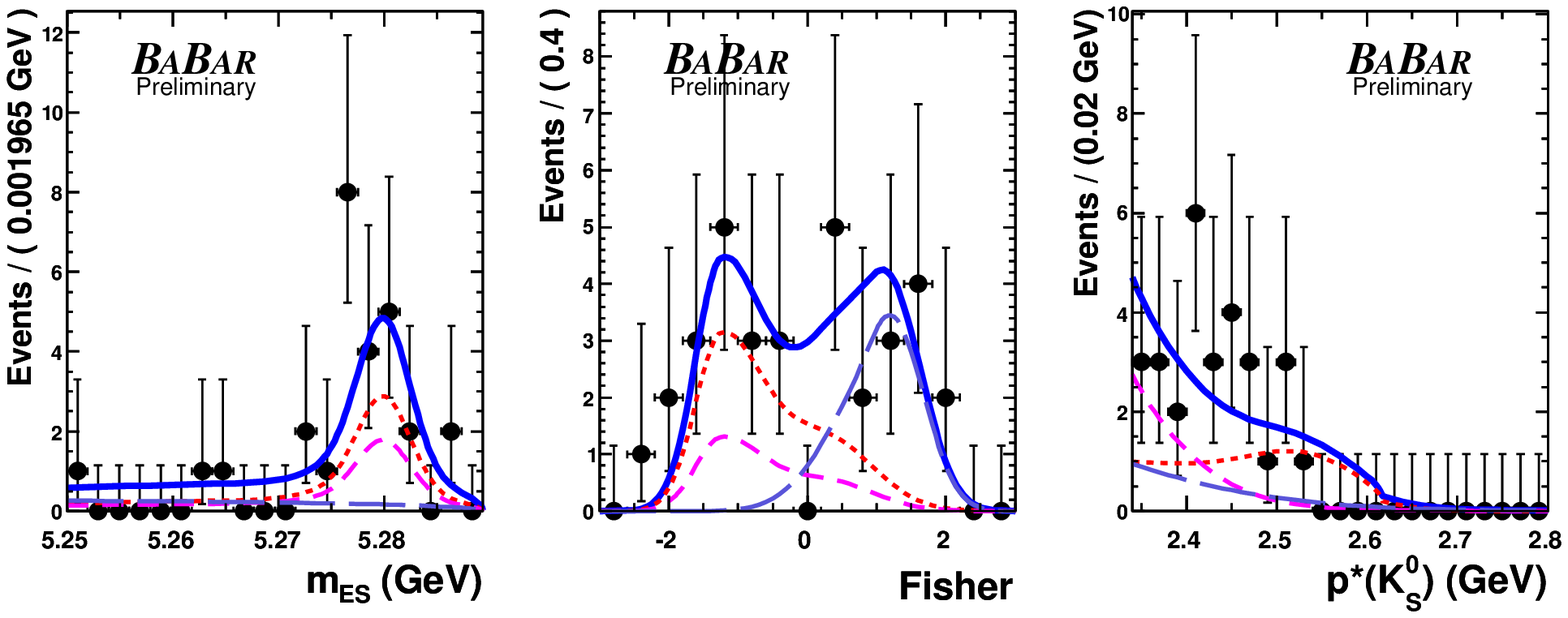} 
\end{center}
\caption{\label{fig:ProjectionPlots}%
Projection plots
for \mes\ (left), \xf\ (center), and \pstarK\ (right)
from the fits to the $\pstarK>2.34$\gev\ samples.
The top plots are for the \BKpX\ decay, and the bottom plots for \BKsX.
The projections are obtained with a cut on the signal likelihood (see text)
retaining about $85\%$ of the \BKpX\ signal events
and $75\%$ of the \BKsX\ signal events.
The points are from the data, the full line shows the full fit,
the dotted line the signal, the short-dashed line the \BB\ background,
and the long-dashed line the \qqbar\ continuum background.}
\end{figure}

The partial branching fractions are calculated as
\begin{equation}
{\cal B}(\BKX,\pstarK>2.34\gev) = \frac{Y_{KX}-Y_b}{\epsilon\cdot N(\breco)},
\end{equation}
where $Y_{KX}$ is the measured yield, $Y_b$ is the fit bias, and
$\epsilon$ is the reconstruction efficiency.
The results of the fits to the $\pstarK>2.34$\gev\ samples and
the quantities used in the determination of the branching
fractions are presented in Table~\ref{tab:results}.
The significance is taken as the square root of the difference
between the value of $-2\ln{\cal L}$ (with additive systematic
uncertainties included) for zero signal and the value at its
minimum.
\begin{table}[!htb]
\caption{Number of events to fit $N_{\rm cand}$, fitted signal yield $Y_{KX}$
in events (ev.), measured bias $Y_b$ (see text), detection efficiency $\epsilon$,
significance~$\cal S$ (with systematic uncertainties included), and measured
partial branching fraction \calB\ for each mode. The first errors are
statistical and the second are systematic. The quantity in parentheses
is the 90\% C.L. upper limit for the branching fraction \brBKzX .}
\begin{center}
\begin{tabular}{lccccccc}
\dbline
Mode	& $N_{\rm cand}$
		& $Y_{KX}$ (ev.)		& $Y_b$ (ev.)
						& $\epsilon$ (\%)
								& $\cal S$ ($\sigma$)
										& \calB\ $(10^{-6})$ \\
\sgline
\BKpX\	& 246	& $58.4^{+10.5}_{-9.7}$	& $2.2$	& $16.1$	& $\signifBKpX$	& $\VpBRBKpX$ \phantom{ ($<\ulpBRBKsX$}	\\
\BKzX	& 76	& $21.1^{+6.5}_{-5.7}$	& $2.8$	& $6.7$		& $\signifBKsX$	& $\VpBRBKsX$ ($<\ulpBRBKsX$)	\\
\dbline
\end{tabular}
\end{center}
\label{tab:results}
\end{table}
In Fig.~\ref{fig:ProjectionPlots} we show projections onto \mes , \xf\
and \pstarK\ of subsamples enriched with a threshold requirement on
the signal likelihood computed without the variable plotted.

\section{SYSTEMATIC STUDIES}
\label{sec:Systematics}
We determine systematic uncertainties affecting the measurement of
the yields, the estimation of the selection efficiencies, and
the measurement of the number of \breco\ candidates. The systematic
uncertainties are summarized in Table~\ref{tab:systematics}.

The signal yield systematic errors arise from the fixed \BB\
yields ($31.0\pm6.8$ events for the charged decay mode and
$13.8\pm4.2$ events for the neutral mode), which are varied
within their uncertainties;
the fit bias correction, for which we assign as systematic error
the quadratic sum of the statistical uncertainty on the correction
and one half of the correction itself;
and the PDF parameter uncertainties, which are left free one by one in
the fit and the variation in signal yield recorded.
The dominant contribution to the PDF parameter uncertainties arises
from the poorly known signal \pstarK\ spectrum. We evaluate this
uncertainty by floating in the fit the relative size and the mean of the Gaussian
used in the description of the signal \pstarK\ PDF.

Uncertainties on the selection efficiencies are dominated
by the statistics of the inclusive \BKX\ Monte Carlo samples.
We also include $0.5\%$ uncertainty per track, $2.1\%$ for
the \KS, and $2.4\%$ for the $K^+$ particle identification criteria.

The uncertainty in the number of fitted \breco\ candidates is
taken from the results of that fit~($5\%$).

\begin{table}[!ht]
\caption{Systematic uncertainties for the \BKpX\ and \BKzX\
decay modes. The multiplicative errors are fractional, and apply
to the efficiency and to the \breco\ counting, while the additive errors
are in units of events and apply to the signal yields.}
\label{tab:systematics}
\begin{center}
\begin{tabular}{l|cc}
\dbline
                                & \BKpX\                & \BKzX         \\
    \sgline
    Multiplicative errors (\%)  & \\
    ~~MC eff                    & $9.4$                 & $16.1$        \\
    ~~Tracking eff/qual         & $0.5$                 & $1.0$         \\
    ~~\KS\ eff                  & --                    & $2.1$         \\
    ~~Kaon PID                  & $2.4$                 & --            \\
    ~~Number \breco\            & $5.0$                 & $5.0$         \\
    \sgline
    Total multiplicative (\%)   & $10.9$                & $17.0$                \\
    \sgline
    Additive errors (events)    & \\
    ~~Fixed \btoc\ yield        & $^{+6.0}_{-5.6}$      & $^{+4.1}_{-3.5}$              \\ 
    ~~PDF parametrization       & $^{+2.6}_{-2.4}$      & $^{+3.9}_{-0.8}$              \\ 
    ~~Fit bias                  & $\pm1.2$              & $\pm1.4$              \\
    \sgline
    Total additive (events)     & $^{+6.6}_{-6.2}$      & $^{+5.8}_{-3.9}$              \\
\dbline
\end{tabular}
\end{center}
\end{table}

\section{CONCLUSIONS}
\label{sec:Summary}
We have presented preliminary results for a study of inclusive charmless
\BKpX\ and \BKzX\ decays, recoiling against fully reconstructed hadronic
$B$ decays from $\UfourS$ decays.
We measure the partial branching fractions for charged and neutral
kaons with momentum above the end-point for \btoc\ backgrounds
($\pstarK>2.34$\gev):
\begin{center}
\begin{tabular}{l}
$\pbrBKpX = (\pBRBKpX ) \times 10^{-6}$, and\\
$\pbrBKzX = (\pBRBKsX ) \times 10^{-6}$ ($<\ulpBRBKsX\times 10^{-6}$ at $90\%$ C.L.).
\end{tabular}
\end{center}
Known exclusive charmless two-body decays, dominated by the
decays $B^+\rightarrow\etapr K^+$ and $B^0\rightarrow\etapr\Kz$,
account for approximately $60\%$ of these branching fractions.
Similar two-body decays with a $K^*$ meson and three-body
decays, such as $B^+\rightarrow K^+K^-K^+$ and
$B^0\rightarrow K^+K^-K^0$, probably account for much of the remainder.

A theoretical model is necessary to extrapolate these results
to the full \pstarK\ spectrum, and ultimately to extract a measurement
of the inclusive \bsg\ branching fraction.
Completing this extrapolation and assigning the theoretical systematic
error from the shape of the \pstarK\ spectrum is the focus of ongoing effort.

\section{ACKNOWLEDGMENTS}
\label{sec:Acknowledgments}

We are grateful for the 
extraordinary contributions of our \pep2\ colleagues in
achieving the excellent luminosity and machine conditions
that have made this work possible.
The success of this project also relies critically on the 
expertise and dedication of the computing organizations that 
support \babar.
The collaborating institutions wish to thank 
SLAC for its support and the kind hospitality extended to them. 
This work is supported by the
US Department of Energy
and National Science Foundation, the
Natural Sciences and Engineering Research Council (Canada),
Institute of High Energy Physics (China), the
Commissariat \`a l'Energie Atomique and
Institut National de Physique Nucl\'eaire et de Physique des Particules
(France), the
Bundesministerium f\"ur Bildung und Forschung and
Deutsche Forschungsgemeinschaft
(Germany), the
Istituto Nazionale di Fisica Nucleare (Italy),
the Foundation for Fundamental Research on Matter (The Netherlands),
the Research Council of Norway, the
Ministry of Science and Technology of the Russian Federation, 
Ministerio de Educaci\'on y Ciencia (Spain), and the
Particle Physics and Astronomy Research Council (United Kingdom). 
Individuals have received support from 
the Marie-Curie IEF program (European Union) and
the A. P. Sloan Foundation.

\end{document}